\documentclass[utf8]{frontiersFPHY} % for Physics and Applied Mathematics and Statistics articles
\setcitestyle{square} % for Physics and Applied Mathematics and Statistics articles
\usepackage{url,hyperref,lineno,microtype,subcaption}
\usepackage[onehalfspacing]{setspace}
\usepackage{amsmath}
\usepackage{booktabs}
\usepackage{multirow}
\usepackage{braket}
\usepackage{caption}
\usepackage{color}

\newcommand{\be}{\begin{equation}}
\newcommand{\ee}{\end{equation}}
\newcommand{\sgmvec}{\ensuremath{\boldsymbol{\sigma}}}
\newcommand{\tauvec}{\ensuremath{\boldsymbol{\tau}}}

\newcommand{\rvec}{{\boldsymbol{r}}}
\newcommand{\rhat}{\hat{\boldsymbol{r}}}

\newcommand{\etavec}{\boldsymbol{\eta}}
\def\keyFont{\fontsize{8}{11}\helveticabold }
\def\firstAuthorLast{Li Muli {et~al.}} %use et al only if is more than 1 author
\def\Authors{Simone Salvatore Li Muli\,$^{1,2,*}$, Sonia Bacca\,$^{1,2}$ and Nir Barnea\,$^{3}$}
\def\Address{$^{1}$Institut f\"ur Kernphysik and PRISMA$^+$ Cluster of Excellence, Johannes Gutenberg Universit\"at, 55128
  Mainz, Germany\\
$^{2}$Helmholtz-Institut Mainz, Johannes Gutenberg Universit\"at Mainz, D-55099 Mainz, Germany\\
$^3$Racah Institute of Physics, The Hebrew University, Jerusalem 91904, Israel
  }
\def\corrAuthor{Simone Salvatore Li Muli}

\def\corrEmail{silimuli@uni-mainz.de}

\begin{document}
\onecolumn
\firstpage{1}

\title[Running Title]{Implementation of local chiral interactions in the hypershperical harmonics formalism} 

\author[\firstAuthorLast ]{\Authors} %This field will be automatically populated
\address{} %This field will be automatically populated
\correspondance{} %This field will be automatically populated

\extraAuth{}% If there are more than 1 corresponding author, comment this line and uncomment the next one.
%\extraAuth{corresponding Author2 \\ Laboratory X2, Institute X2, Department X2, Organization X2, Street X2, City X2 , State XX2 (only USA, Canada and Australia), Zip Code2, X2 Country X2, email2@uni2.edu}

\maketitle

\begin{abstract}
%%% Leave the Abstract empty if your article does not require one, please see the Summary Table for full details.
With the goal of using chiral interactions at various orders to explore  properties of the few-body nuclear systems, we write the recently developed local chiral interactions as spherical irreducible tensors and implement them in the hyperspherical harmonics expansion method. We devote particular attention to three-body forces at next-to-next-to leading order, which play an important role in reproducing experimental data. We check our implementation by  benchmarking the  ground-state properties  of $^3$H, $^3$He and $^4$He against the available Monte Carlo calculations. We then confirm their order-by-order truncation error estimates and further investigate uncertainties  in the charge radii obtained by using the precise muonic atom data for single-nucleon radii. Having local chiral Hamiltonians at various orders implemented in our hyperspherical harmonics suites of codes opens up the possibility to test such interactions on other light-nuclei properties, such as electromagnetic reactions.

\tiny
 \keyFont{ \section{Keywords:} nuclear interactions, hyperspherical harmonics, light nuclei, ab-initio theory, chiral perturbation theory}
\end{abstract}

\section{Introduction}
%Importance of chiral EFT in nuclear physics. Order-by-order to assess uncertainties. Progress from few to many-body nuclei. Need to look at new observables ($^4$He monopole and muonic atoms).Importance of benchmarking before calculating new observables.

In 1935 the seminal idea of Yukawa \cite{yuk35} laid the foundation to the theory of the nuclear forces. His one-pion exchange term is nowadays known as an important contribution to the interaction among nuclei in the long-distance range and is implemented in many nuclear interaction models. In the mid 1990s the first high-precision nucleon-nucleon (NN) potentials able to reproduce at the same time the deuteron properties, the proton-proton and the proton-neutron scattering data were released. Some notable examples of these interactions are the Argonne $v18$ (AV18) \cite{wir95}, the Nijmegen (Nijm93) \cite{sto93} and the charge-dependent Bonn (CD-Bonn) \cite{mac01}. The subsequent development of three-nucleon (3N) interactions, see for instance Refs.~\cite{pud95,pie01}, improved the description of the $A>2$ nuclear dynamics, initiating a successful theoretical campaign of nuclear structure and reaction predictions, see e.g., Refs.~\cite{Leidemann:2012hr,BaccaPastore,Noemi} and references therein. Despite the great success of the phenomenological interactions, there are still open questions to address, including the difficulty of providing solid uncertainty quantifications in the modeling of the forces, the lack of connection between the NN and 3N interactions and the missing direct link to quantum chromodynamic (QCD), the fundamental theory of the strong force.

An important step forward to address these issues was made when the concept of effective field theory (EFT) was introduced and applied to low-energy QCD. As suggested by Weinberg \cite{wei79,wei90,wei91,wei92}, the low-energy nuclear dynamic can be described by a Lagrangian written in terms of pions and nucleons fields and consistent with all the commonly accepted symmetries of QCD, including the (explicitly and spontaneously broken) chiral symmetry which strongly constrains the pion dynamics. The proposed Lagrangian contains an infinite number of terms and a systematic expansion must be introduced to make the theory applicable. Following Weinberg's proposal, in the early 2000s modern versions of chiral-inspired nuclear interactions were released by many groups -- for a compilation of results see for instance Refs.~\cite{epe09,mac11,epe20} and references therein -- each interaction being different by the truncation order of the chiral expansion, by the inclusion or exclusion of the $\Delta(1232)$-isobar, by the fitting procedure or by the regularization scheme used. Given that these interactions are derived in field theories written in momentum space, they are highly non-local. One of the consequence is that they are   difficult to implement in some of the few- and many-body techniques which are developed in coordinate-space representation.

In recent years a new chiral-inspired set of nuclear interactions at the next-to-next-to-leading order (N2LO) has become available~\cite{gez14,lyn16,lyn17}. These interactions have a series of interesting properties which make them a promising framework for future nuclear computations. These interactions are completely written in coordinate space and contain only one non-local operator. Furthermore the NN and 3N terms are  regularized consistently, namely the same regulator form and cut-off is used. Here, these interactions are written for the first time as product of irreducible tensors under space rotations, a required step for the implementation into the hyperspherical harmonics formalism. Using the method of hyperspherical harmonics, we perform benchmark tests in light-nuclear systems, where we compare to available results from the Green's function Monte Carlo (GFMC) and the auxiliary field diffusion Monte Carlo (AFDMC) methods.

% Finally, we do an order-by-order analysis of the electric-dipole polarizability of the $^4$He nucleus in terms of the chiral order. This analysis represents one of the greatest advantages of using EFT-inspired interactions over phenomenological models, since it allows for a solid evaluation of the nuclear uncertainties.

%The authors intention is to extend the work performed in this paper by looking at observables that have never been considered for testing the nuclear interactions before. In particular the recent advances of the CREMA (charge radii experiments in muonic atoms) collaborations \cite{poh10,ant13,poh16} suggest that muonic atoms might represent an ideal ground for benchmarking the nuclear forces with unprecedented precision, improving in particular the so far poorly constrained 3N forces.

This review paper is summarized as follows. In Section 2, we briefly overview the formulation of the hyperspherical harmonics method in coordinate-space representation.
In Section 3, we present the maximally-local chiral interactions developed in Ref.~\cite{lyn17} and rewrite the 3N force  as products of irreducible tensors under space rotations. In Section 4, we show our benchmark results for $^3$H, $^3$He and $^4$He and we discuss uncertainties. 
% and the order-by-order analysis of the electric-dipole polarizability of $^4$He in the chiral expansion.
 Finally, Section 5 is reserved for the conclusive remarks and the overview of future prospects. 

\section{Hyperspherical harmonics}
The hyperspherical harmonic method was firstly introduced in 1935 by Zernike and Brinkman \cite{zer35}, reintroduced later in the 60's by Delves \cite{del59}, Simonov \cite{sim66}, Zickendraht \cite{zic65} and Smith \cite{smi60} and it is extensively applied nowadays to the study of few-body systems. For recent reviews with applications to nuclear physics we refer the reader to the following references~\cite{kie08, mar20}. In this work, the hyperspherical harmonic functions are constructed to form irreducible representations of the $SO(3)$ group of space rotations, the $O(N)$ group of dynamical rotations in the space spanned by the $N$ Jacobi vectors, and the $S_A$ permutation group of the $A$-particle system. The method is briefly reviewed in this section, the formalism introduced follows closely  Refs.~\cite{bar97,bac05}.

We consider a system of $A$ identical nucleons, the Jacobi coordinates $\{\etavec_i\}$ are commonly introduced in order to separate the internal degrees of freedom from the center of mass. There are several ways to construct the set of $N=A-1$ Jacobi coordinates out of the $A$ coordinate vectors $\{\rvec_i\}$ of the nucleons. One commonly used definition for the relative Jacobi vectors is
\begin{equation}
\etavec_{j-1}=\sqrt{\frac{j-1}{j}}\Bigl(\rvec_j - \frac{1}{j-1}\sum_{i=1}^{j-1}\rvec_i\Bigr); \quad k=2,...,A.
\end{equation}
From a given choice of Jacobi coordinates, the hyperspherical coordinates $\{\rho_N, \varphi(N), \Omega(N)\}$ can be introduced. In this notation,  $\rho_N$ is the hyper-radius,
 $\Omega(N)\equiv \{\Omega_1,...,\Omega_N\}$ where $\Omega_j=(\theta_j,\phi_j)$ gathers the angular coordinates of the Jacobi vectors ,
and $\varphi(N)\equiv \{\varphi_2,...,\varphi_N\}$ is a set of hyper-angles. 

The hyper-radial coordinates $\rho_1,...,\rho_N$ and the hyper-angular coordinates $\varphi_2,...,\varphi_N$ are constructed recursively. The transformation law for the first two Jacobi coordinates is

\begin{equation}
\begin{aligned}
\label{hc}
\eta_1=\rho_1=\rho_2\cos \varphi_2, \\
\eta_2=\rho_2\sin \varphi_2.
\end{aligned}
\end{equation} 

Assuming that we already know the hyper-radial coordinates $\rho_1,...,\rho_{j-1}$ and the hyper-angular coordinates $\varphi_2,...,\varphi_{j-1}$ the transformation law for $\rho_j$ and $\varphi_j$ reads in analogy to Eq.~(\ref{hc}) as

\begin{equation}
\begin{aligned}
\rho_{j-1}=\rho_j\cos \varphi_j, \\
\eta_j=\rho_j\sin \varphi_j.
\end{aligned}
\end{equation} 

The internal kinetic energy operator for the A-body system is given by the $3N$-dimensional Laplace operator $\Delta(N)$. In terms of the hyperspherical coordinates it is written as

\begin{equation}
\Delta(N) = \Delta_\rho - \frac{1}{\rho^2}\hat{K}^2_N (\varphi(N),\Omega_1,...,\Omega_N)
\end{equation}
where the hyper-radial part is 
\begin{equation}
\Delta_\rho = \frac{\partial^2}{\partial \rho^2} + \frac{3N-1}{\rho}\frac{\partial}{\partial \rho},
\end{equation}
while $\hat{K}_N^2$ is the grand-angular momentum operator whose eigenfunctions are known as the hyperspherical harmonics. 

Denoting $\hat{l}_j$ as the angular momentum operator related to $\etavec_j$, and $\hat{L}^2_j$ and $\hat{M}_{j}$ as the total orbital angular momentum operator and $z$-projection  of the system identified by the first $j$ Jacobi coordinates, it is possible to define the grand-angular momentum operator  $\hat{K}_N^2$ of the system recursively  in terms of $\hat{K}^2_{N-1}$ and $\hat{l}_N$ as \cite{efr72}

\begin{equation}
\hat{K}_N^2= - \frac{\partial^2}{\partial \varphi_N^2} + \frac{3N-6- (3N-2) \cos (2\varphi_N)}{\sin (2\varphi_N)}\frac{\partial}{\partial \varphi_N} + \frac{1}{\cos^2 \varphi_N} \hat{K}_{N-1}^2+ \frac{1}{\sin^2 \varphi_N} \hat{l}^2_N
\end{equation}
where  $\hat{K}_1^2=\hat{l}^2_1$. 

The operators $\hat{K}_N^2,...,\hat{K}_{2}^2$, $\hat{L}^2_N,...,\hat{L}^2_2$, $\hat{l}^2_N,...,\hat{l}^2_1$ and $\hat{M}_{N}$ commute with each others. As a consequence, it is possible to label hyperspherical states using the set of $3N-1$ quantum numbers $\{K\}\equiv\{K_N,...,K_2,L_N,...,L_2,l_N,...,l_1,M_{N}\}$. The hyperspherical harmonics functions $\mathcal{Y}_{\{K_N\}}$ are the eigenfunctions of the grand-angular momentum operator with eigenvalues $K_N(K_N+3N-2)$. The explicit expression for the resulting hyperspherical harmonics functions is given by \cite{fab83}

\begin{equation}
\begin{split}
\mathcal{Y}_{\{K_N\}} &= \Biggl[ \sum_{m_1,...,m_N} C_{l_1m_1,l_2m_2}^{L_2M_2} C_{L_2M_2,l_3m_3}^{L_3M_3}\times ... \times C_{L_{N-1}M_{N-1},l_{N}m_{N}}^{L_NM_N} \prod_{j=1}^N Y_{l_jm_j}(\Omega_j) \Biggr]\times \\
& \times \Biggl[ \prod_{j=2}^N \mathcal{N}_j (\sin \varphi_j)^{l_j} (\cos \varphi_j)^{K_{j-1}} P_{n_j}^{\bigl[l_j+\frac{1}{2}, K_{j-1}+\frac{(3j-5)}{2}\bigr]}(\cos 2\varphi_j) \Biggr],
\end{split}
\end{equation}
where $C^{LM}_{l_im_i,l_jm_j}$ are the Clebsch-Gordan coefficients, $Y_{l_jm_j}(\Omega_j)$ are the spherical harmonics associated with $\etavec_j$ and
\begin{equation}
\begin{split}
&\mathcal{N}_j= \Biggl[ \frac{(3K_j+3j-2) n_j! \Gamma(n_j+K_{j-1}+l_j+\frac{3j-2}{2})}{\Gamma(n_j+l_j+\frac{3}{2}) \Gamma(n_j+K_{j-1}+\frac{3j-3}{2})} \Biggr]^\frac{1}{2}
\end{split}
\end{equation}
is a normalization constant with
$2n_j=K_j - K_{j-1}- l_j$.

In our formulation of the hyperspherical harmonics method we construct hyper-angular functions that form irreducible tensors under the $SO(3)$ group of spatial rotations, the $O(N)$ group of kinematic rotations and the $S_A$ group of permutations of the $A$ nucleons. These symmetry-adapted hyperspherical harmonics, $\mathcal{Y}_{[K_N]}$, are uniquely identified by the set of quantum numbers $[K_N]\equiv\{K_N,L_N,M_N,\boldsymbol{\lambda}_N,\alpha_N\,Y_A,\beta_A\}$. For the current purposes, it is enough to specify that $\boldsymbol{\lambda}_N$ identifies the irreducible representation of $O(N)$, $Y_A$ is the Yamanouchi symbol which specifies the irreducible representations of the group-subgroup chain $S_1 \subset ... \subset S_A$ presented by the appropriate Young diagrams $\Gamma_1,..., \Gamma_A$, while $\alpha_N$ and $\beta_A$ are additional quantum numbers needed to remove further degeneracies. The $O(N)$ and $S_A$ symmetry-adapted hyperspherical harmonics $\mathcal{Y}_{[K_N]}$ are constructed recursively. Assuming that $\mathcal{Y}_{[K_{N-1}]}$ have been already constructed, the $N$th Jacobi coordinate is then coupled to this system, so that a state with total angular momentum $L_N$ and grand-angular momentum $K_N$ is formed, let us call this state $\mathcal{Y}_{[K_{N-1}],K_NL_NM_N}$. Note that $\mathcal{Y}_{[K_{N-1}],K_NL_NM_N}$ is a irreducible tensor under $O(N-1)$ and $S_{A-1}$ but not under $O(N)$ and $S_A$. The states $\mathcal{Y}_{[K_N]}$ are obtained as linear combinations of the states $\mathcal{Y}_{[K_{N-1}],K_NL_NM_N}$, where the coefficients of the linear combinations are labeled as $\Bigl[ \bigl( K_{N-1}, L_{N-1}, \boldsymbol{\lambda}_{N-1}, \alpha_{N-1}; l_N \bigr) K_N L_N |\} K_N L_N \boldsymbol{\lambda}_N \alpha_N \Bigr]$, $\Bigl[ \bigl( \boldsymbol{\lambda}_{N-1} \Gamma_{N} \beta_N\bigr) \boldsymbol{\lambda}_N |\} \boldsymbol{\lambda}_N \Gamma_A \beta_A \Bigr]$ and are known as hyperspherical orthogonal group parentage coefficients (hsopcs) and orthogonal group coefficients of fractional parentage (ocfps) respectively. 

The full expression of the symmetry-adapted hyperspherical harmonics reads
\begin{equation}
\begin{split}
\mathcal{Y}_{[K_N]} &= \sum_{\boldsymbol{\lambda}_{N-1}\beta_{N}}^{} \Bigl[ \bigl( \boldsymbol{\lambda}_{N-1} \Gamma_{N} \beta_N\bigr) \boldsymbol{\lambda}_N |\} \boldsymbol{\lambda}_N \Gamma_A \beta_A \Bigr] \times \\
&\times \sum_{K_{N-1},L_{N-1},\alpha_{N-1},l_N} \Bigl[ \bigl( K_{N-1}, L_{N-1}, \boldsymbol{\lambda}_{N-1}, \alpha_{N-1}; l_N \bigr) K_N L_N |\} K_N L_N \boldsymbol{\lambda}_N \alpha_N \Bigr] \times \\
& \times \mathcal{Y}_{[K_{N-1}],K_NL_NM_N}.
\end{split}
\end{equation}

Nucleons posses also spin and isospin degrees of freedom. Because the nuclear Hamiltonian is rotationally invariant, nuclear states have the total angular momentum $J$ as good quantum number. Furthermore, isospin is an approximate symmetry for the nuclear interaction with the consequence that the total isospin $T$ of a nuclear state is a conserved quantum number. For these reasons we couple the symmetry-adapted hyperspherical harmonics to the $S_A$ symmetry-adapted spin-isospin wavefunction $\chi$ of the $A$-nucleon system

\begin{equation}
H_{(K_N)} = \sum_{Y_N} \frac{\Lambda_{\Gamma_A, Y_N}}{\sqrt{|\Gamma_A|}} \sum_{M_NS_z} C_{L_NM_N,SS_z}^{JJ_z} \mathcal{Y}_{[K_N]}~ \chi_{[S_A]}\,.
\end{equation}
Here $(K_N)\equiv \{K_NL_NS_NJ_NJ_{N_z}\boldsymbol{\lambda}_N\alpha_N^{ST}Y_A\beta_A\}$, $[S_A]\equiv \{SS_zTT_zY_A\alpha_A^{ST}\}$,  $\Lambda_{\Gamma_A, Y_N}$ is a phase factor, and $|\Gamma_A|$ is the dimension of the irreducible representation $\Gamma_A$. 

Analogously to what has been done with the hyperspherical harmonics, the spin-isospin wavefunctions are constructed recursively. Assuming that the symmetry-adapted wavefunction $\chi_{[S_{j-1}]}$ have been obtained, the construction of the $\chi_{[S_{j}]}$ is done by first coupling $\chi_{[S_{j-1}]}$ to the spin-isospin wavefunction of the $j$th nucleon, let us call this state $\chi_{[S_{j-1}],S_jT_j}$, and then taking linear combinations of $\chi_{[S_{j-1}],S_jT_j}$ using the coefficients of fractional parentage labeled as $\Bigl[S_{j-1}  S_j T_{j-1} T_j\Gamma_{j-1}\alpha_{j-1}^{ST}|\}S_jT_j\Gamma_j\alpha_j^{ST}\Bigr]$. Namely the full expression for $\chi_{[S_j]}$ reads

\begin{equation}
\chi_{[S_j]}= \sum_{S_{j-1}T_{j-1}\alpha_{j-1}^{ST}}\Bigl[S_{j-1}S_jT_{j-1}T_j\Gamma_{j-1}\alpha_{j-1}^{ST}|\}S_jT_j\Gamma_j\alpha_j^{ST}\Bigr] \chi_{[S_{j-1}],S_jT_j}\,. 
\end{equation}

We are finally able to expand the nuclear wavefunction in terms of hyperspherical harmonics. In practice, the expansion is performed up to a maximal value of the grand-angular quantum number $K_{max}$ as

\begin{equation}
\Psi = \sum_{(K_N)} \mathcal{R}_{(K_N)}(\rho_N) H_{(K_N)}(\Omega_N)\,.
\end{equation}
When we insert this wavefunction into the Schrödinger equation, an eigenvalues equation is obtained for the hyper-radial wavefunction $\mathcal{R}_{(K_N)}$, the eigenvalue equation is then solved by expanding the hyper-radial wavefunction in terms of an orthogonal set of functions. In this work the set is taken as the generalized Laguerre polynomials $L^v_n(\rho_N)$. Again, the model space is truncated to a given maximum number of Laguerre polynomials $n_{max}$ 
\begin{equation}
\mathcal{R}_{(K_N)} = \sum_{n=0}^{n_{max}} C^n_{(K_{N})} L^v_n(\rho_N).
\end{equation}
With the introduction of this further model space, the resulting eigenvalue equation is solved with direct diagonalization routines, or with the Lanczos method when the model space is too big for a direct diagonalization.
In essence, the hyperspherical harmonics method is a powerful technique that allows for an exact solution of the Schrödinger equation for few-body systems. In the limit where $n_{max}\rightarrow \infty$ and $K_{max}\rightarrow \infty$ the solution correspond to the exact solution to the Schrödinger equation. While we observe that good convergence can be reached with $n_{max}\le 50$, the convergence in terms of $K_{max}$  will be carefully investigated. 
The uncertainty coming from the truncation of the model space, in particular of $K_{max}$, can be estimated by looking at the convergence pattern of the observables of interest, for instance the binding energy and the radius. As a consequence, the method is an excellent candidate for uncertainty quantifications in nuclear physics, with the possibility of performing tests over commonly accepted nuclear Hamiltonians or making precise predictions for few-nucleon systems. Because the formulation we present here is developed in coordinate space, the method benefits from having local
forces, such as the AV18 potential. While one can formulate hyperspherical harmonics also in momentum space~\cite{HHp}, the goal of this paper is to work in coordinate space and implement local-chiral interactions.
To further improve the convergence with respect to the model space, we make use of the effective interaction hyperspherical harmonics (EIHH) method. The interested reader can find  more details on this approach in Ref.~\cite{EIHH}, and also in the more recent review~\cite{muon_review}.

\section{Nuclear Hamiltonians}

Nuclear physics is mainly formulated in the framework of non-relativistic quantum mechanics. The relevant degrees of freedom are represented by the nucleons, whose interactions are remnants of the color forces among the quarks. In this picture, the nucleus is a compound object of $A$ non-relativistic nucleons and the dynamic of the system is specified by the nuclear Hamiltonian operator

\begin{equation}
\hat{H}=\hat{T}+\hat{V}+\hat{W} + ... =\sum^A_{i=1} \hat{T}_i+\sum^A_{i>j=1}\hat{V}_{ij}+\sum^A_{i>j>k=1} \hat{W}_{ijk} +... \,,
\end{equation}
where $\hat{T}$ is the sum of the non-relativistic kinetic energy operators of the individual nucleons, $\hat{V}$ is a sum of NN interactions and $\hat{W}$ is a sum of 3N interactions. The dots stand for higher order forces not explicitly included in this work.

Our goal is to solve the Sch\"odinger equation
\be
\hat{H} \ket{\Psi} = E \ket{\Psi}
\ee
and  when working with antisymmetrized wavefunctions, the expectation values of the NN and 3N terms become
\begin{eqnarray}
\label{eq:expval}
\braket{\Psi|\hat{V}|\Psi}&=&\frac{A(A-1)}{2} \braket{\Psi|\hat{V}_{12}|\Psi},\\
\nonumber
\braket{\Psi|\hat{W}|\Psi}&=&\frac{A(A-1)(A-2)}{6} \braket{\Psi|\hat{W}_{123}|\Psi}\,,
\end{eqnarray}
 where only the first two (or three) particles are involved\footnote{This property will be used later when we will write explicitly the form of the nuclear forces between (among) two (three) particles.}.

In the modern theory of nuclear forces, interactions are derived from the chiral effective field theory (ChEFT). In this theory, proposed first by Weinberg \cite{wei79,wei90,wei91,wei92}, the chiral Lagrangian is constructed in terms of pion and nucleon fields and is consistent with the commonly accepted symmetries of QCD, including the explicitly and spontaneously broken chiral symmetry. This effective Lagrangian has infinitely many terms, therefore one needs to introduce an ordering scheme  to render the theory predictive. 

In ChEFT, the terms in the chiral Lagrangian are analyzed counting powers of a small external momentum over the large scale : $(Q/\Lambda_{\chi})^\nu$, where $Q$ stands for an external momentum or a pion mass and $\Lambda_{\chi}$ is the chiral symmetry breaking scale, whose value is approximately given by the mass of the $\rho$-meson $\Lambda_{\chi}\sim m_\rho=770$ MeV. Determining systematically the power of $\nu$ has become known as power counting. The lowest possible value of $\nu$ is conventionally referred to as the leading order (LO), the second lowest is the next-to-leading order (NLO), the third lowest is the next-to-next-to leading order (N2LO) and so on. While there are many proposed power counting schemes \cite{kap98,kap98nov,nog05,pav06,lon12,van94}, in this work we adopt the Weinberg power counting, which makes use of naive dimensional analysis~\cite{wei90,wei91}. 

Given that ChEFT is naturally formulated in momentum space, the derived nuclear interactions are strongly non-local,
which is a disadvantage for methods that are formulated in coordinate space. However, it has been recently found that it is possible to construct maximally local chiral interactions by regularizing in coordinate space and exploiting Fierz ambiguities to remove non-localities in the short-distance interactions \cite{gez14,lyn16,lyn17}.

The local chiral NN forces are composed by contact $(ct)$ terms and pion-exchange $(\pi)$ terms so that
the interaction  between  particle 1 and 2 can be written as
\be
V_{12}= V^{ct}_{12} + V^\pi_{12}\,.
\ee

 When working with totally anti-symmetric systems, it is possible to exploit Fierz ambiguities for removing the non-local operators contributing to the contact NN interactions. This means that  the interactions
 can be chosen to have the following operator structure \cite{gez14} at LO
\be
V^{ct,\rm LO}_{12}= \bigl(C_{\rm S} + C_{\rm T} \sgmvec_1 \cdot \sgmvec_2 \bigr) \delta (\rvec_{12}),
\ee
where $\rvec_{12}$ is the relative distance between nucleon $1$ and nucleon $2$, $\sgmvec_{1/2}$ are the vector-spin Pauli matrices operating in the space of the first/second nucleon\footnote{Even though they are operators in spin space, we do not use the hat in our notation, as they are vectors, whose components are operators.} and $\delta$ is the delta function. 

At NLO, the following new terms enter
\be
\begin{split}
V^{ct,\rm NLO}_{12} = &
% V^{ct,\rm LO}(\rvec_{12}) 
-\bigl(C_{\rm 1} + C_{\rm 2} \tauvec_1 \cdot \tauvec_2 \bigr) \Delta \delta (\rvec_{12}) \cr
&-\bigl(C_{\rm 3} + C_{\rm 4} \tauvec_1 \cdot \tauvec_2 \bigr) \sgmvec_1 \cdot \sgmvec_2 \Delta \delta (\rvec_{12}) \cr
&+ \frac{C_{5}}{2}\frac{\partial_{r_{12}} \delta(\rvec_{12})}{r_{12}} \textbf{L}\cdot \textbf{S} + \bigl( C_6 +C_7 \tauvec_1 \cdot \tauvec_2 \bigr) \cr
& \times \Biggl[ \bigl(\sgmvec_1 \cdot \rvec_{12}\bigr) \bigl(\sgmvec_2 \cdot \rvec_{12}\bigr) \Bigl[ \frac{\partial_{r_{12}} \delta (\rvec_{12})}{r_{12}} - \partial_{r_{12}}^2 \delta (\rvec_{12}) \Bigr] - \sgmvec_1 \cdot \sgmvec_2 \frac{\partial_{r_{12}}\delta (\rvec_{12})}{r_{12}}\Biggr],
\end{split}
\ee
where  $\tauvec_{1/2}$ are the vector-isospin Pauli matrices, $\Delta$ is the Laplace operator, $\textbf{L}$ and $\textbf{S}$ are the total orbital angular momentum and spin operator in the two-body system represented by the two interacting nucleons\footnote{We drop the hat from vectors whose components are operators.}, and the $\delta$-function will have to be regularized. The $\{C_i\}$ are a set of low energy constants (LECs). The term proportional to the LEC $C_5$ is the only non-local operator appearing in this maximally local chiral interaction.

 Following Refs.~\cite{gez14,pia20}, all the pion-exchange interactions up to N2LO can be written in a complete local form as
\begin{align}
V^\pi_{12} &= V_{\rm C}(\rvec_{12}) + W_{\rm C}(\rvec_{12}) \tauvec_1 \cdot \tauvec_2 \cr
&+ \bigl[V_{\rm S}(\rvec_{12}) + W_{\rm S}(\rvec_{12}) \tauvec_1 \cdot \tauvec_2 \bigr]\sgmvec_1 \cdot \sgmvec_2 \cr
&+ \bigl[V_{\rm T}(\rvec_{12}) + W_{\rm T}(\rvec_{12}) \tauvec_1 \cdot \tauvec_2 \bigr] S_{12}\,,
\end{align}
where, $S_{12}$ is the well known tensor operator, defined as 
\be
   S_{12} = 3(\sgmvec_1\cdot\rhat_{12})(\sgmvec_2\cdot\rhat_{12})-(\sgmvec_1\cdot\sgmvec_2)\,,
\ee
where $\rhat_{12}$ is the unitary vector related to the relative distance $\rvec_{12}$.
The local functions $V_{\rm C}(\rvec_{12})$, $W_{\rm C}(\rvec_{12})$, $V_{\rm S}(\rvec_{12})$, $W_{\rm S}(\rvec_{12})$, $V_{\rm T}(\rvec_{12})$ and $W_{\rm T}(\rvec_{12})$ have dependencies on the axial-vector coupling constant of the nucleon $g_A$, on the pion decay constant $\rm{F}_\pi$ and on the pion mass $\rm{m}_\pi$. These functions are evaluated at each order in ChEFT (LO, NLO and N2LO) and details can be found in Ref.~\cite{pia20}.
% and at N2LO they also depend on the coupling constants denoted with $c_1,c_2,c_3$ and $c_4$, which are fit in the pion nucleon sector and are .
 In Ref.~\cite{lyn17} pion loops are regularized using the spectral-function regularization (SFR) with  an ultraviolet cut-off $\tilde{\Lambda}=1$ GeV and we follow this prescription. 

The local chiral NN interactions up to N2LO are already written or can be written with minimal modifications as irreducible tensors under space rotations. Thus, they can be easily implemented in the hyperspherical harmonics formalism in coordinate space.  In fact, they have pretty much the same structure as the Argonne potential AV8'~\cite{av8p}. 
The same does not apply to 3N interactions.

% In this work the spin, isospin and angular structure of the 3N operators are coupled together as irreducible tensors under SO(3), then we show our benchmark tests on light nuclei where we compare with monte-carlo computations, finally the convergence of the electric-dipole polarizability is analyzed for the first time using the N2LO 2N+3N local chiral interaction.     

\begin{figure}[h!]
\begin{center}
\includegraphics[scale=0.15] {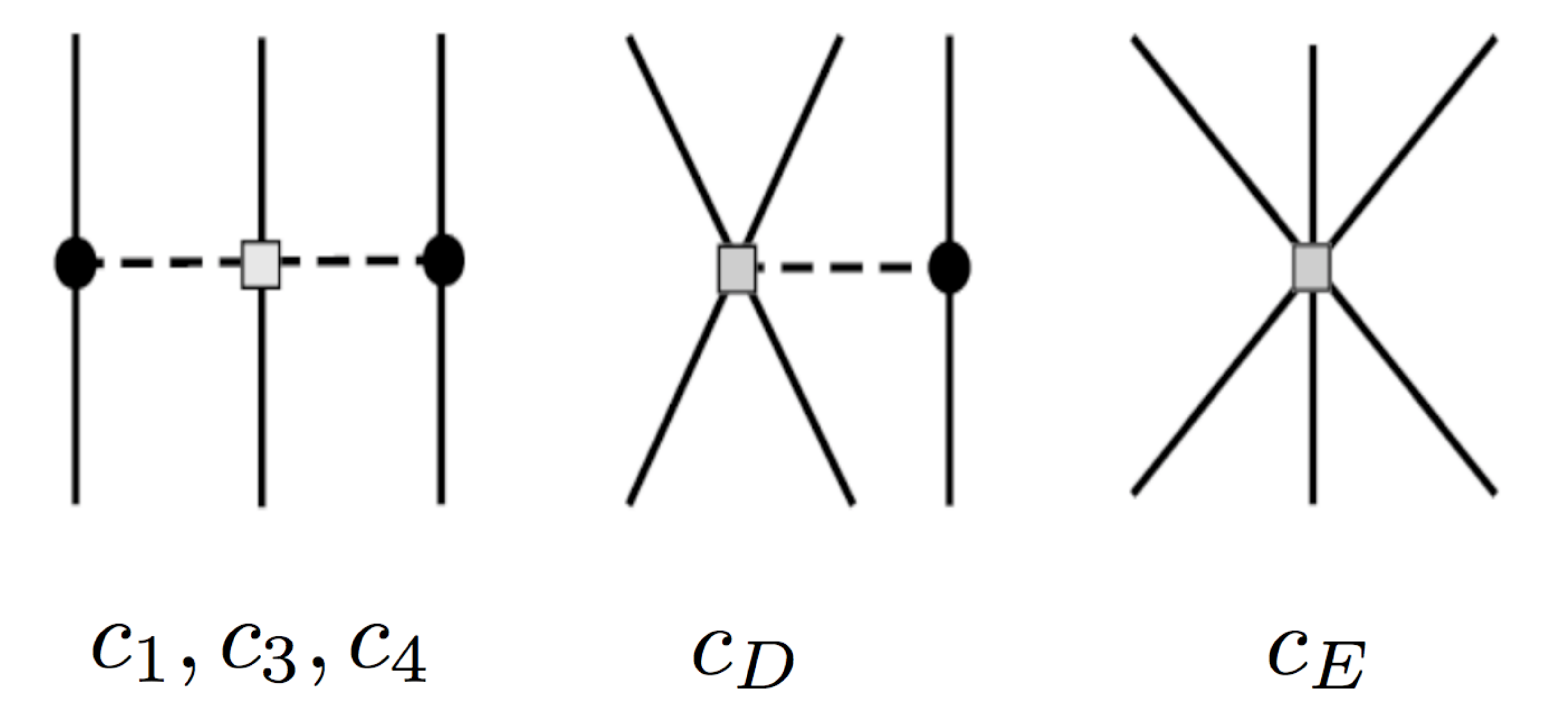}
\caption{Feynman diagrams of the chiral 3N force at N2LO, from the left to the right: $2\pi$-term, $1\pi$-term and $ct$-term.} \label{diag}
\end{center}
\end{figure}

Three-body interactions arise at NLO in Weinberg power counting. However, at this order their contribution is canceled out. The first non-zero contributions start at N2LO. The 3N force at this order is composed of a two-pion ($2\pi$) exchange, a one-pion ($1\pi$) exchange and a 3N contact ($ct$) interaction, see Fig.~\ref{diag}. On the one hand, the $2\pi$-term comes with the LECs $c_1$, $c_3$ and $c_4$ that already appear at the subleading two-pion-exchange interaction at the NN level at the same chiral order which highlights the consistency of the NN and 3N interactions in ChEFT. On the other hand, the one-pion exchange and the 3N contact diagrams introduce two new LECs, $c_D$ and $c_E$, which must be fitted on $A\geq 3$ observables.

With respect to Ref.~\cite{lyn17}, here the 3N interaction is written for a given triplet of nucleons, since at the end we use the fact that the wavefunction is anti-symmetric to compute the expectations values as in Eq.~(\ref{eq:expval}). The 3N interaction reads
\be
 W_{123}=\sum_{\text{cyc}}W_{1,23}=\sum_{\text{cyc}}\Bigl[ W^{2\pi, c_1}_{1,23}+W^{2\pi, c_3}_{1,23}+W^{2\pi, c_4}_{1,23}
              +W^{1\pi, c_D}_{1,23} +W^{ct,c_E}_{1,23}\Bigr] \;,
\ee
where the sum runs over the cyclic permutations of the particle triplet and the notation has the intention to highlight the symmetry of the interaction over the exchange of particles 2 and 3.   Each term is denoted with a label that includes the associated LEC. 
 
The $2\pi$ exchange terms are given by
\begin{eqnarray}
\label{2pi}
W^{2\pi,c_1}_{1,23} &=& A U_{12}Y_{12}U_{13}Y_{13}
            (\tauvec_2\cdot\tauvec_3 )
            (\sgmvec_2\cdot\rhat_{12})(\sgmvec_3\cdot\rhat_{13}),
\cr
W^{2\pi,c_3}_{1,23} &=& B 
   \{\tauvec_1\cdot\tauvec_2,\tauvec_1\cdot\tauvec_3\}\{\chi_{12},\chi_{13}\},
\cr 
W^{2\pi,c_4}_{1,23} &=& -C 
   [\tauvec_1\cdot\tauvec_2,\tauvec_1\cdot\tauvec_3][\chi_{12},\chi_{13}]\,,
\end{eqnarray}
where the coupling constants are
$A=c_1\frac{g_A^2m_\pi^4(\hbar c)^2}{16\pi^2 F_{\pi}^4},\;
  B=c_3\frac{g_A^2m_{\pi}^4 (\hbar c)^2}{1152\pi^2 F_{\pi}^4},\;$ and 
 $ C=c_4\frac{g_A^2m_{\pi}^4 (\hbar c)^2}{2304\pi^2 F_{\pi}^4}$.
The $2\pi$-terms include the following functions
\begin{eqnarray}
  Y_{12}&=Y(r_{12})=\frac{e^{-m_\pi r_{12}}}{r_{12}},\\
\nonumber
  U_{12}&=U(r_{12})=1+\frac{1}{m_\pi r_{12}},
\end{eqnarray}
with analogous expressions for $Y_{13}$ and $U_{13}$.
The operator $\chi_{12}$ (and analogously $\chi_{13}$) is defined as  
\be
  \chi_{12}=X_{12}-\frac{4\pi}{m_{\pi}^2}\delta_{12}\sgmvec_1\cdot\sgmvec_2        
          =T_{12}S_{12}+\tilde{Y}_{12}\sgmvec_1\cdot\sgmvec_2
\ee
with
\begin{eqnarray}
  X_{12}&=&T_{12}S_{12}+Y_{12}\sgmvec_1\cdot\sgmvec_2,\\
\nonumber
 \tilde{Y}_{12} &=& Y_{12}-\frac{4\pi}{m_{\pi}^2}\delta_{12}\,,\\
\nonumber
  \delta_{12}&=&\delta_{r_0}(r_{12})
             =\frac{1}{\frac{4\pi}{n}\Gamma(\frac{3}{n})r_0^3}e^{-(r_{12}/r_0)^n}.
\end{eqnarray}	
In the last expression, $r_0$ is the cut-off and following Refs.~\cite{lyn17,lyn16} $n$ is taken to be equal to 4.% $\Gamma(3/n)=\Gamma(3/4)$.

For the $1\pi$-interaction terms there are two options 
\begin{eqnarray}
\label{1pi}
W^{1\pi,c_D1}_{1,23} &=& D (\tauvec_2\cdot\tauvec_3 )
    [X_{23}(\rvec_{12})\delta_{13}+X_{23}(\rvec_{13})\delta_{12}
      -\frac{8\pi}{m^2_\pi}\delta_{12}\delta_{13}\sgmvec_2\cdot\sgmvec_3]\\
\nonumber
&{\rm and}&\\
\nonumber
W^{1\pi,c_D2}_{1,23} &=& D (\tauvec_2\cdot\tauvec_3 )\chi_{23}(\delta_{12}+\delta_{13})\,,
\end{eqnarray}
with
 $D=c_D\frac{g_A m_{\pi}^2(\hbar c)^4}{96 \pi \Lambda_{\chi} F_{\pi}^4}$.
While the difference between the two is due to regulator artifacts, in this work only the second choice is implemented, namely $W^{1\pi,c_D2}_{1,23}$. 

For the contact term there are different options on the operator structure, which come from different choices in the Fierz rearrangement. In this work the following two are considered
\begin{eqnarray}
\label{ct}
W^{ct,c_E\tau}_{1,23} &=& E ({\tauvec}_2\cdot{\tauvec}_3 ) \delta(r_{12})\delta(r_{13})\\ 
\nonumber
&{\rm and}&\\
\nonumber
W^{ct,c_E1}_{1,23} &=& E \delta(r_{12})\delta(r_{13}) 
\end{eqnarray}
with
$ E=c_E\frac{(\hbar c)^6}{\Lambda_{\chi}F_{\pi}^4}.$
%It is worth noticing that $\Lambda_\chi, m_{\pi}, F_\pi$ are given in [MeV], and $c_1,c_3,c_4$
%in $[\text{MeV}^{-1}]$.
%The units of $A,B,C$ are $[\text{MeV}\times\text{fm}^2]$,
%the units of $D$ are $[\text{MeV}\times\text{fm}^4]$, and the units
%of $E$ are $[\text{MeV}\times\text{fm}^6]$. 
\begin{table}[h!t]
\centering
\begin{tabular}{ccccccc}
\toprule
\multirow{2}{*}{3N force} & r$_0$  & $c_E$ & $c_D$ & $c_1$ & $c_3$ & $c_4$\\
 & [fm] &  &  & [GeV$^{-1}$] & [GeV$^{-1}$] & [GeV$^{-1}$]\\
\hline
& & & & & & \\
\multirow{2}{*}{N2LO (D2, E$\tau$)} & 1.0 & -0.63 & 0.0 &-0.81 & -3.20 & 3.40 \\
 & 1.2 & 0.085 & 3.5 & -0.81 & -3.20 & 3.40 \\
\bottomrule
\end{tabular}
\caption{Fit values for the couplings $c_D$ and $c_E$ for different choices of 3N cut-offs as reported in \cite{lyn16,lyn17}. The constants $c_{1,3,4}$ are tuned in the pion-nucleon sector, see Ref.~\cite{mac11}.}
\label{tab:couplings}
\end{table}

The value of all LECs entering the 3N forces at N2LO are shown in Table~\ref{tab:couplings}.
 In Refs.~\cite{lyn16,lyn17} $c_D$ and $c_E$ have been fitted in order to reproduce the $^4$He binding energy and the $n$-$\alpha$ $P$-wave phase shift.

%%%%\subsection{Conventions and Notation}

\subsection{Three-nucleon forces as spherical tensors}

The above expressions for the 3N force are not written 
 in terms of irreducible spherical tensors, so that they can not be implemented directly  into the hyperspherical formalism. In this section we address this point and write the interaction in terms of irreducible spherical tensors, both in coordinate-spin space and in isospin space.  

For convenience, we denote the general spin space $\Sigma_{ij}^\lambda$, $\Sigma_{ij,k}^{\lambda,\Lambda}$  and configuration space $ X_{ij}^{\lambda}$, $X_{ij,ij}^{(\lambda,\lambda')\Lambda}$ irreducible tensor operators as
\begin{eqnarray}
\Sigma_{ij}^{\lambda} &=& [\sgmvec_i \times \sgmvec_j]^{\lambda},
\cr
\Sigma_{ij,k}^{\lambda, \Lambda} &=& 
        \left[[\sgmvec_k \times 
              [\sgmvec_i \times \sgmvec_j]^{\lambda}
        \right]^{\Lambda},
\cr
X_{ij}^{\lambda} &=& [\hat{\rvec}_{1i} \times \hat{\rvec}_{1j} ]^{\lambda}, \cr
X_{ij,ij}^{(\lambda,\lambda')\Lambda} 
     &=& \left[ [\hat{\rvec}_{1i} \times \hat{\rvec}_{1j} ]^{\lambda}
       \times [\hat{\rvec}_{1i} \times \hat{\rvec}_{1j} ]^{\lambda'} \right]^{\Lambda},
\end{eqnarray}
 where $i,j,k$ are generic particle indexes and
 $\hat{\rvec}_{1i}$ is the rank 1 normalized spherical tensor associated to the relative distance between particle 1 and particle $i$. With the notation $ [\rhat_{1i}\times \rhat_{1j} ]^{\lambda}$ we intend the two rank-one coordinate space tensors coupled into a rank-$\lambda$ tensor, and analogously for $[\sgmvec_i \times \sgmvec_j]^{\lambda}$ and $ [\tauvec_i\times\tauvec_j]^\lambda$ in spin and isospin space, respectively. Furthermore, we define
\be
X^{\lambda}(\rvec_{ij},\rvec_{ij}) = [\hat{\rvec}_{ij} \times \hat{\rvec}_{ij} ]^{\lambda}\,,
\ee
where  $\hat{\rvec}_{ij}$ is the rank 1 normalized spherical tensor associated to the relative distance between particles $i$ and $j$.

At this point, after rearranging the couplings with  a few Racah algebra steps and by using the previously introduced notation, one can  rewrite the
 3$N$ interactions of Eqs.~(\ref{2pi}),(\ref{1pi}),(\ref{ct}) in terms of irreducible tensors in isospin space and in the coupled spin-configuration space.

%\paragraph*{The $2\pi-c_1$ term}
The $2\pi$-exchange term depending on $c_1$ becomes
\begin{align}
W^{2\pi, c_1}_{1,23} &= A U_{12}Y_{12}U_{13}Y_{13}
            (\tauvec_2\cdot\tauvec_3 )
            (\sgmvec_2\cdot\rhat_{12})(\sgmvec_3\cdot\rhat_{13})
\cr &=
           -\sqrt{3} A [\tauvec_2\times\tauvec_3]^0 F_{UU}
         \left(\Sigma_{23}^0\cdot X^0_{23}
              -\Sigma_{23}^1\cdot X^1_{23}
              +\Sigma_{23}^2\cdot X^2_{23}\right)\,,
\end{align}
%\paragraph*{The $2\pi-c_3$ term}
 the $2\pi$-exchange term that depends on $c_3$ becomes
\begin{align}
W^{2\pi, c_3}_{1,23} &= B 
              \{\tauvec_1\cdot\tauvec_2,\tauvec_1\cdot\tauvec_3\}\{\chi_{12},\chi_{13}\}
\cr &=
     -2\sqrt{3}B[\tauvec_2\times\tauvec_3]^0
     \Big( 
\cr & \hspace{3em}
      \Sigma_{23}^0\cdot (
              +F_{TT} X^{0}_{23}
              -\frac{1}{\sqrt{3}}(3F_{YY}+F_{TY}+F_{YT})
              )
\cr & \hspace{2em} +
      \Sigma_{23}^1\cdot (
               -F_{TT} X^{1}_{23})
\cr & \hspace{2em} +
      \Sigma_{23}^2\cdot (
              +F_{TT} X^{2}_{23}
              +F_{TY} X^{2}_{22}+F_{YT} X^{2}_{33}
              )
     \Big)\,,
\end{align}
while the term  that depends on $c_4$ can be expressed as
%\subsubsection{The $2\pi-c_4$ term}
\begin{align}
W^{2\pi, c_4}_{1,23} &= -C 
              [\tauvec_1\cdot\tauvec_2,\tauvec_1\cdot\tauvec_3][\chi_{12},\chi_{13}]
\cr
&= 4\sqrt{3} C [\tauvec_1\times[\tauvec_2\times\tauvec_3]^1]^0 \Big[
\cr & \hspace{3em}
      \Sigma_{23,1}^{1,0}\cdot \Big(-F_{TT}X^{(1,1)0}_{23,23}
           -\frac{1}{\sqrt 3}(3F_{YY}+F_{TY}+F_{YT})\Big)
\cr & \hspace{2em} +
      \Sigma_{23,1}^{0,1}\cdot \Big(+F_{TT}X^{(1,0)1}_{23,23}\Big)
\cr & \hspace{2em} +
      \Sigma_{23,1}^{2,1}\cdot \Big(+F_{TT}X^{(1,2)1}_{23,23}\Big)
\cr & \hspace{2em} +
      \Sigma_{23,1}^{1,2}\cdot \Big(-F_{TT}X^{(1,1)2}_{23,23}
           -{\frac{1}{2}}(F_{TY}X_{22}^2+F_{YT}X_{33}^2)\Big)
\cr & \hspace{2em} +
      \Sigma_{23,1}^{2,2}\cdot \Big(-F_{TT}X^{(1,2)2}_{23,23}
           -{\frac{\sqrt 3}{2}}(F_{TY}X_{22}^2-F_{YT}X_{33}^2)\Big)
\cr & \hspace{2em} +
      \Sigma_{23,1}^{2,3}\cdot \Big(+F_{TT}X^{(1,2)3}_{23,23}\Big)
      \Big]\,.
\end{align}
To write the above expression in a compact form, we have introduced the following definitions
\begin{eqnarray}
  F_{UU}&=&    U_{12}Y_{12}U_{13}Y_{13}, \cr
  F_{TT}&=& 18 T_{12}T_{13} ,\cr
  F_{YY}&=&  2 (\tilde{Y}_{12}-T_{12})(\tilde{Y}_{13}-T_{13}), \cr
  F_{TY}&=&  6 T_{12}(\tilde{Y}_{13}-T_{13}), \cr
  F_{YT}&=&  6 (\tilde{Y}_{12}-T_{12})T_{13} \,.
\end{eqnarray}

%\subsubsection{The $D$ term}
The $1\pi$-exchange contribution  takes the following form
\begin{align}
W^{1\pi, c_D2}_{1,23} &= D (\tauvec_2\cdot\tauvec_3 )\chi_{23}(\delta_{12}+\delta_{13})
\cr
&=
         -\sqrt{3}D[\tauvec_2\times\tauvec_3 ]^0\Big[
\cr & \hspace{3em}
           \Sigma_{23}^0\cdot \Big(
              -\sqrt{3}(\delta_{12}+\delta_{13})\tilde{Y}_{23}\Big)
 \cr &\hspace{2em} +
           \Sigma_{23}^2\cdot \Big(
           3 (\delta_{12}+\delta_{13})T_{23} X^2(\rhat_{23},\rhat_{23})\Big)
         \Big]\,,
\end{align}
%\subsubsection{The $E$ term}
while the contact terms  become
\begin{eqnarray}
W^{ct,c_E\tau}_{1,23} &=& E ({\tauvec}_2\cdot{\tauvec}_3 ) \delta_{12}\delta_{13} =
     -\sqrt{3} E [\tauvec_2\times\tauvec_3 ]^0 \delta_{12}\delta_{13} \\
\nonumber
&{\rm and}&\\
\nonumber
W^{E1}_{1,23} &=& E \delta_{12}\delta_{13}\,.
\end{eqnarray}

We have implemented these expressions in our hypershperical harmonics codes. Since the interaction is now written in terms of irreducible tensors, the spin and isospin matrix elements can be computed analytically. For the calculation of the spatial matrix elements one can reduce the six-dimensional integration in the two Jacobi coordinates to a two-dimensional numerical quadrature, as explained in details in Ref.~\cite{Barnea3NF}.
 Below we present the benchmark results we obtained with these local-chiral forces on few-body systems such as $^3$H, $^3$He and $^4$He.

\section{Results}
In this section we show the benchmark tests of the maximally-local-chiral interactions using the EIHH method. We compute ground-state energies and charge radii in three- and four-nucleon systems and compare to two Monte Carlo methods, namely the GFMC and AFDMC methods.

In the computations of nuclear charge radii, we use
\begin{equation}
\label{rad_eq}
\braket{r^2_{\text{c}}} = \braket{r^2_{\text{pt}}} + r^2_{\text{p}} + \frac{A-Z}{Z} r^2_{\text{n}} + \frac{3\hbar^2}{4m^2_pc^2},
\end{equation}
were $\sqrt{\braket{r^2_{\text{pt}}}}$ is the calculated point-proton radius, $r_{\text{p}}=0.8751(61)$ fm \cite{moh16} is the root-mean-square (rms) charge radius of the proton, $r^2_{\text{n}}=-0.1161(22)$ fm$^2$ \cite{moh16} is the squared charge radius of the neutron, and $Z$ is the number of protons in the nucleus. The last term is the Darwin-Foldy correction to the proton-charge radius \cite{fri97} which depends on the proton mass $m_p$. We neglect the spin-orbit relativistic contribution, since it is negligible in $s$-shell nuclei~\cite{Ong}, as well as meson exchange currents.

 Keeping in mind that the goal of this work is to benchmark our expressions for the 3N forces at N2LO  by comparing to the Monte Carlo results, we have used the same numerical value for $r_p$ and $r_n$ as in Ref.~\cite{lyn17}, which follows the CODATA-2014 recommendations~\cite{moh16}. Hence, in a first stage we will not be using the more modern results for $r_{p/n}$ from Refs.~\cite{ant13,fil20}.

A few words addressing the estimation of the numerical uncertainties are in line. As  already  said, the EIHH method allows for an exact solution of the Schrödinger equation, the computed wavefunction converges to the true eigenfunction of the Hamiltonian operator in the limit of infinite model space. The model space is mostly given by the maximal number, $n_{\text{max}}$, of Laguerre polynomials and the choice of the maximal value of the grand-angular momentum quantum number, $K_{\text{max}}$, in the construction of the hyperspherical harmonics functions. It has been practically found that beyond a value $n_{\text{max}}=50$, the expectation values are negligibly modified. The convergence in terms of $K_{\text{max}}$ is more delicate, so that in order to estimate the uncertainty coming from the truncation of the model space, we analyze the converging pattern at increasing values of $K_{\text{max}}$.  
%In all the final results shown in this paper the maximal value of $K_{\text{max}}$ we used is 22.

To quantify our numerical uncertainty we proceed as follows.  Denoting with $O(K_{\text{max}})$  the expectation value of an observable $\hat{O}$ computed by setting a given maximal value of the grand-angular momentum quantum number, $K_{\text{max}}$, in the wavefunction, our uncertainty in this observable is estimated by
\begin{equation}
\label{uq}
\delta(O) = |O(K_{\rm max})-O(K_{\rm max}-2)| + |O(K_{\rm max}-2)-O(K_{\rm max}-4)| + \delta_{\text{res}},
\end{equation}
where $\delta_{\text{res}}$ is the residual uncertainty (not due to the $K_{max}$ behavior) obtained by varying: the number of radial grid points (from 70 to 90), the maximal values of the  angular momentum in the construction of the two-body effective interaction (from 60 to 120) and the maximal number of  three-body  angular momentum (from 5/2 to 7/2) in the partial wave expansion of the 3N force. 

First, we address and discuss the benchmark of the interactions at LO and NLO, so to have a clean test on the NN interactions. Then we move to the N2LO, where the three-body forces are included.

\subsection{Benchmarks at LO and NLO}

We study the maximally-local chiral interactions for two different regulator cut-offs, indicated by $r_0$, namely exploring the two possibilities of $r_0=1.0$ fm and $r_0=1.2$ fm. The latter gives rise to a softer interaction compared to the first one. For the benchmarks at LO and NLO, the $^4$He nucleus is used as a testing ground. We compute point-proton charge radii, $\sqrt{\braket{\text{r}^2_{\text{pt}}}}$, and ground-state energies, $E_0$, for the two different cut-off choices at increasing values of the grand-angular momentum quantum number and compare to the GFMC calculations.

\begin{table}[h!t]
\centering
\begin{tabular}{cccc c ccc}
\toprule
 & \multicolumn{3}{c}{LO} & & \multicolumn{3}{c}{NLO}\\
\cline{2-4}
\cline{6-8}

\multirow{2}{*}{} & Cut-off & E$_0$ & $\sqrt{\braket{\text{r}^2_{\text{pt}}}}$ & & Cut-off & E$_0$ & $\sqrt{\braket{\text{r}^2_{\text{pt}}}}$ \\ & [fm] & [MeV] & [fm] & & [fm] & [MeV] & [fm] \\
\cline{2-4}
\cline{6-8}
& & & & & & &\\
\multirow{2}{*}{EIHH} & 1.0 & -42.830(6) & 1.0370(3) & & 1.0 & -21.55(4) & 1.575(1) \\ & 1.2 & -46.6054(7) & 1.01765(4) & & 1.2 & -22.974(6) & 1.5278(6) \\
& & & & & & &\\
\multirow{2}{*}{GFMC} & 1.0 & -42.83(1) & 1.02(1)  & & 1.0 & -21.56(1) & 1.57(1) \\ & 1.2 & -46.62(1) & 1.00(1) & & 1.2 & -22.94(6) & 1.53(1) \\
& & & & & & &\\
Nature & & -28.29566 & 1.46(1) & &  & -28.29566 & 1.46(1) \\
\bottomrule
\end{tabular}
\caption{Ground-state energies and point-proton radii for the $^4$He nuclear system at LO and NLO computed with the EIHH method. For comparison we report the GFMC results and the experimental values taken from Ref.~\cite{lyn14,DataBase}.}
\label{tab:4HeLONLO}
\end{table}
 The final results are shown in Table~\ref{tab:4HeLONLO}, where the uncertainty is computed as explained above using Eq.~(\ref{uq}) with $K_{\rm max}=22$. An extended table with all the various $K_{max}$ can be found in the Supplementary Material. We observe that as we enlarge the model space a nice converging pattern is obtained and our final EIHH results agree with the GFMC calculations within uncertainties. By looking at the converging pattern of the studied observables as the model space is increased (see Supplementary Material), we clearly observe that the interaction with $r_0=1.2$ fm is much softer than the other, since the relative observables converge with a smaller model space. Finally, it is to note that, as shown in Table~\ref{tab:4HeLONLO},
 the LO and NLO results do not reproduce the measured values, but the discrepancy decreases in going from LO to NLO. 

\subsection{Benchmarks at N2LO}

We now turn to the benchmark at the next order.
At N2LO we have the first appearance of 3N forces, so this will serve as a check of our irreducible tensor representation. The 3N interaction involves two new LECs, $c_D$ and $c_E$, coming from the $1\pi$-term and from the $ct$-term of the 3N forces, respectively, that can not be fitted in the NN sector. In Ref.~\cite{lyn16} these couplings have been fitted to reproduce the $^4$He binding energy and the $n$-$\alpha$ scattering $P$-wave phase shift, for which
the values reported in Table~\ref{tab:couplings} were obtained. We use the same values in this work, as our goal is to perform a benchmark. In particular, here we implement only the (D2, E$\tau$)  3N interactions, which we chose since the E$\tau$ term has a more general isospin structure. Different choices of the 3N contact term have been shown to lead to different saturation properties in neutron matter~\cite{lyn16}.

\begin{figure}[h!]
\begin{center}
\includegraphics[scale=0.7] {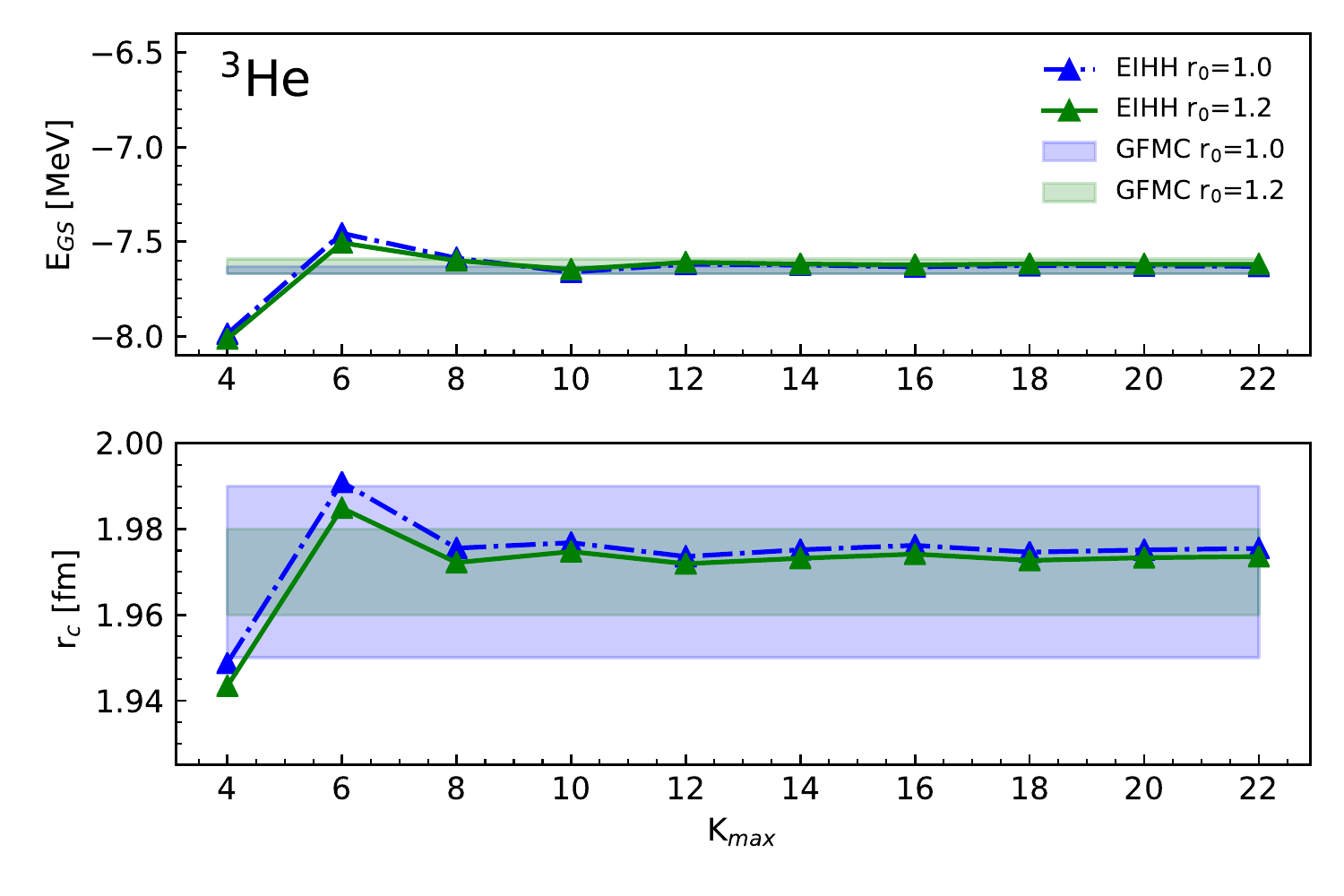}
\end{center}
\caption{The ground-state energy and the charge radius of the nuclear $^3$He system as a function of the grand-angular momentum quantum number $K_\text{max}$. The green and blue error-bands are the GFMC results with the relative statistical uncertainty.}
\label{fig:3He}
\end{figure}

\begin{figure}[h!]
\begin{center}
\includegraphics[scale=0.7] {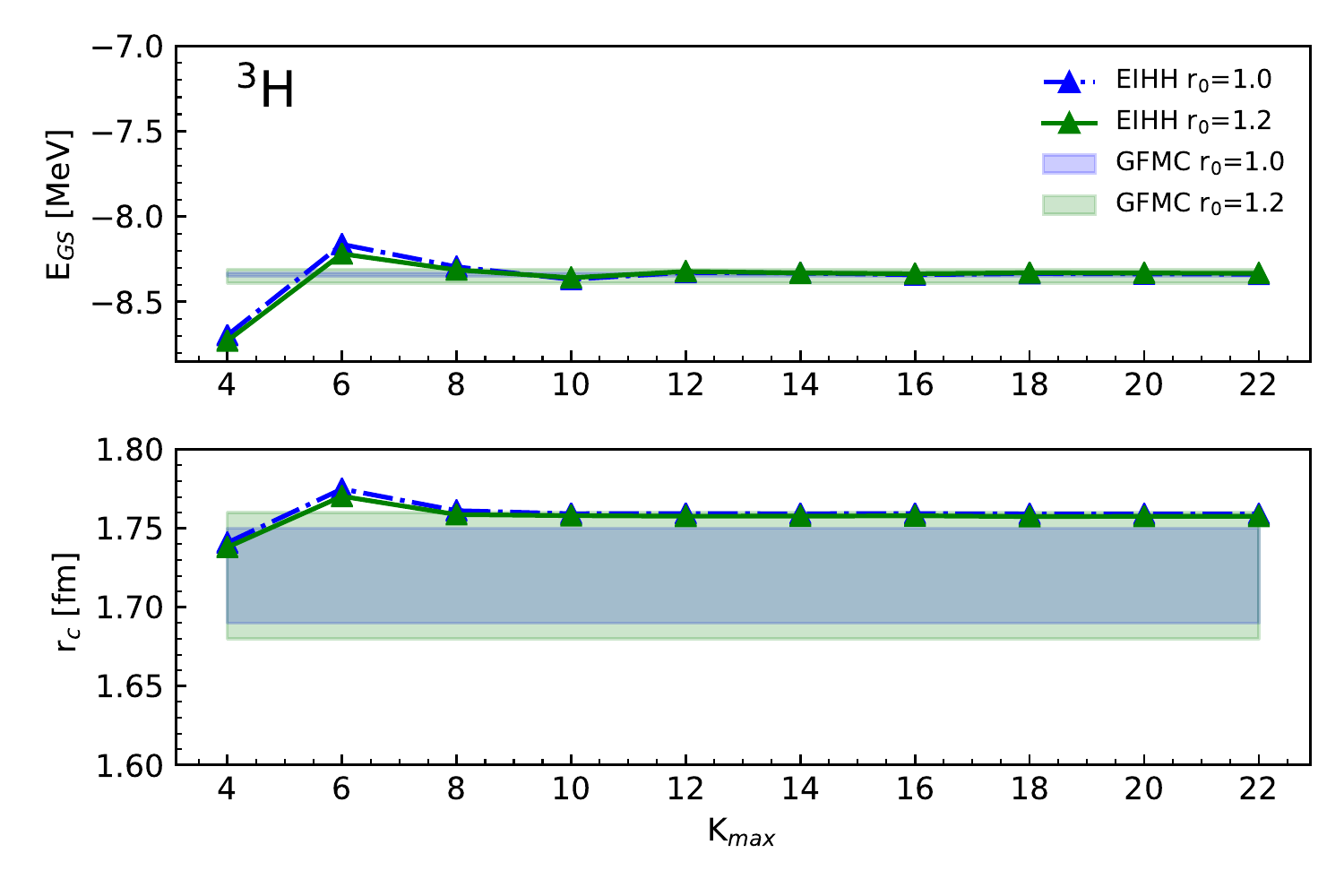}
\end{center}
\caption{The ground-state energy and the charge radius of the nuclear $^3$H system as a function of the grand-angular momentum quantum number $K_\text{max}$. The green and blue error-bands are the GFMC results with the relative statistical uncertainty.}
\label{fig:3H}
\end{figure}

\begin{figure}[h!]
\begin{center}
\includegraphics[scale=0.7] {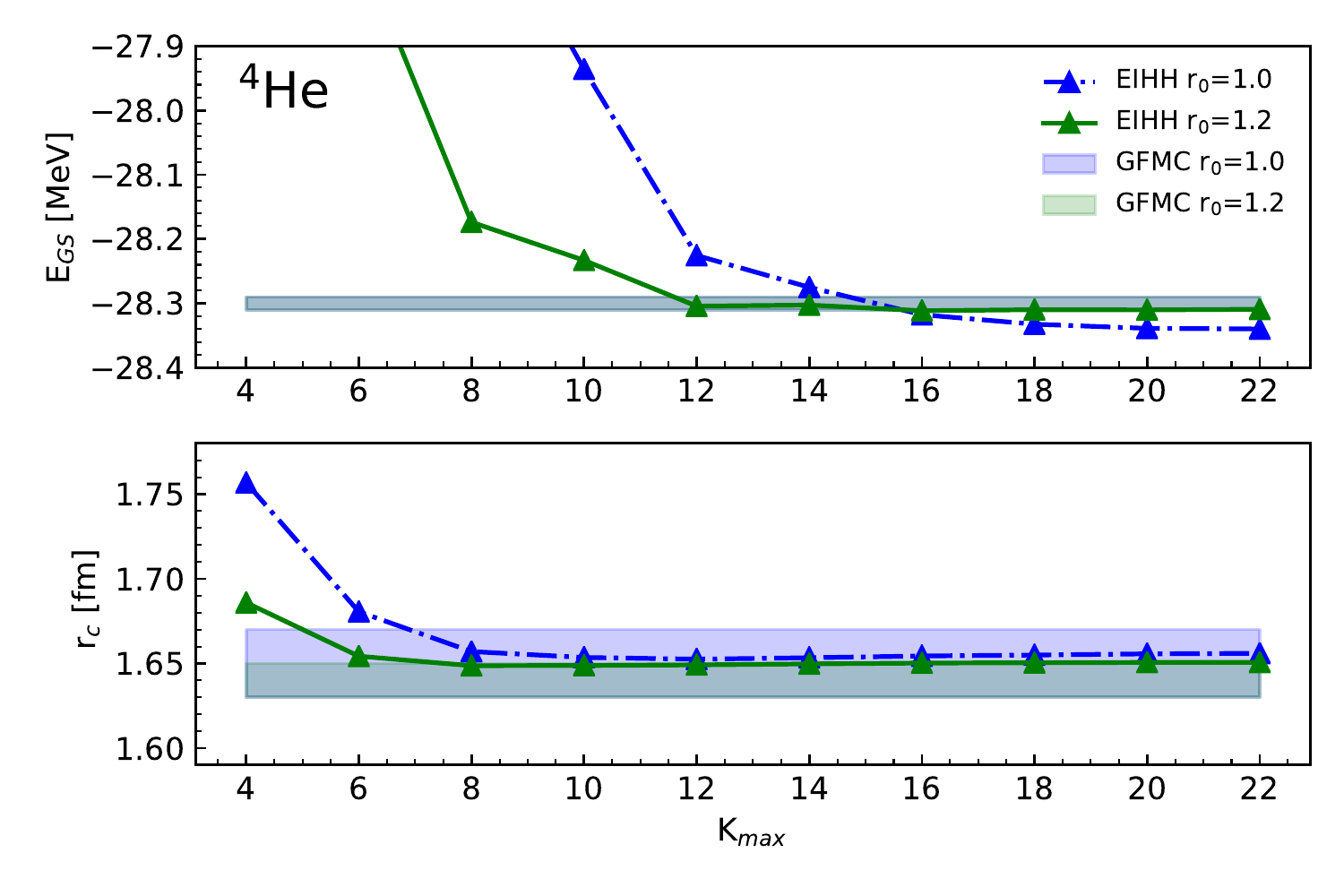}
\end{center}
\caption{The ground state energy and the charge radius of the nuclear $^4$He system as a function of the grand-angular momentum quantum number $K_\text{max}$. The green and blue error-bands are the GFMC results with the relative statistical uncertainty.}
\label{fig:4He}
\end{figure}

As a testing ground for our N2LO Hamiltonian expressed in terms of spherical tensors outlined in the previous section, we study the three-body $^3$He and $^3$H and the four-body $^4$He nuclear systems. We compute ground-state energies, $E_0$, and charge radii, $\sqrt{\braket{\text{r}^2_{\text{c}}}}$, for the two different cut-off choices $r_0=1$ and 1.2 fm and carefully study the convergence at increasing  $K_{\text{max}}$ values. A complete table of our data is shown in the Supplementary Material. The $K_{\rm max}$ convergence  is also explicitly shown in a graphical manner in Fig.~\ref{fig:3He}, Fig.~\ref{fig:3H}, and Fig.~\ref{fig:4He}, where a comparison to the GFMC method is made.

 As it can be seen from Fig.~\ref{fig:3He} and \ref{fig:3H}, the EIHH method is in excellent agreement with the GFMC  computations for the three-body nuclei, for both the ground-state energies and the charge radii. The typical non-monotonic convergence patter of the EIHH method is observed, and a very good convergence is reached already at $K_{max}=12$.
This shows that these forces are softer than the AV18 potential, but harder than the low-q interactions~\cite{bacca_vlowk}.

For the $^4$He nucleus shown in Figure~\ref{fig:4He}, we obtain a very nice agreement with the GFMC method for the cut-off value $r_0=1.2$ fm, while for the cut-off $r_0=1.0$ fm, we perfectly reproduce the charge radius, but we observe 
 a small deviation for the ground-state energy with respect to the GFMC.

\begin{table}[h!t]
\centering
\begin{tabular}{cc ccc c cc cc}
\toprule
 & & \multicolumn{2}{c}{$^3$He} & & \multicolumn{2}{c}{$^3$H} & & \multicolumn{2}{c}{$^4$He}\\
\cline{3-4}
\cline{6-7}
\cline{9-10}

\multirow{2}{*}{} & Cut-off & E$_0$ & $\sqrt{\braket{\text{r}^2_{\text{c}}}}$ & & E$_0$ & $\sqrt{\braket{\text{r}^2_{\text{c}}}}$ & & E$_0$ & $\sqrt{\braket{\text{r}^2_{\text{c}}}}$ \\ & [fm] & [MeV] & [fm] & & [MeV] & [fm] & & [MeV] & [fm] \\
\cline{3-4}
\cline{6-7}
\cline{9-10}
& & & & & & & & & \\
\multirow{2}{*}{EIHH} & 1.0 & -7.630(6) & 1.976(7) & & -8.338(5) & 1.759(6) & & -28.34(5) & 1.656(6) \\ & 1.2 & -7.619(4) & 1.974(5) & & -8.332(3) & 1.758(5) & & -28.31(2) & 1.651(4) \\
& & & & & & & & & \\
\multirow{2}{*}{GFMC} & 1.0 & -7.65(2) & 1.97(2) & & -8.34(1) & 1.72(3) & & -28.30(1) & 1.65(2) \\ & 1.2 & -7.63(4) & 1.97(1) & & -8.35(4) & 1.72(4) & & -28.30(1) & 1.64(1) \\
& & & & & & & & & \\
\multirow{2}{*}{AFDMC} & 1.0 & -7.55(8) & 1.96(2) & & -8.33(7) & 1.72(2) & & -27.64(13) & 1.68(2) \\ & 1.2 & -7.64(4) & 1.95(5) & & -8.27(5) & 1.73(2) & & -28.37(8) & 1.65(1)\\
& & & & & & & & & \\
Nature & & -7.718043(2)   & 1.973(14) & &  -8.481798(2)& 1.759(36) & &-28.29566 & 1.681(4)\\
\bottomrule
\end{tabular}
\caption{Ground-state energies and charge radii for the nuclear $^3$He, $^3$H and $^4$He systems at N2LO in the chiral expansion computed with the EIHH, GFMC and AFDMC method. For the EIHH results, we report the estimation of the uncertainty coming from the truncation of the model space, the errors of the GFMC and AFDMC are statistical. Experimental values are from Ref.~\cite{Sick2014,PURCELL20101,ANGELI201369,DataBase}.}
\label{tab:overv}
\end{table}

 Our final EIHH  results with uncertainties quantified as explained above  using Eq.~(\ref{uq}) with $K_{\rm max}=22$  are shown in Table~\ref{tab:overv}
in comparisons with the GFMC, AFDMC and the experimental data.
 We note that the small difference found for $^4$He ground-state energy is just at the level of $0.03$ MeV in the central values and is non-significant when the full uncertainty of the EIHH method is considered.
Similar kind of sub-percentage differences between EIHH and GFMC were also observed in other benchmarks~\cite{Zemach_bench} and can be found at this level of precision. It is to note that the cut-off $r_0=1.0$ fm leads to a harder force, where in fact, quite a large discrepancy is  seen also between the  GFMC and the AFDMC computations. 
Therefore, we do not think that this difference is significant and we consider all these results to constitute a successful benchmark of our implementation of 3N forces.

As can be seen in Table~\ref{tab:overv}, at N2LO a much improved agreement  with experiment is obtained. In fact, if one compares the experimental binding energies to the LO and NLO calculations in Table~\ref{tab:4HeLONLO} one observes that these low orders overbind (LO) or underbind (NLO) the few-body nuclei, while 
at N2LO nice agreement is observed.
This is expected for $^4$He, given that 
 3N forces are fit to reproduce the $^4$He binding energy, however a better agreement is also found for $^3$He and $^3$H due to the strong correlation between the three- and four-body binding energy. Interestingly, a nice converging pattern is  also found for the nuclear charge radii.

From a careful look at  Table~\ref{tab:overv}, one can appreciate that our EIHH calculations are more precise than the GFMC and AFDMC results in the three-nucleon sector and that our numerical uncertainty is comparable to the experimental uncertainties for the radii.
While this may be an advantage of our method, it is important to note that the error bars quoted in this table do not include the uncertainties 
coming from the ChEFT expansion, so they do not constitute the full 
uncertainty of the theory.

We conclude this section with a further investigation on the charge radii of light-nuclear systems. In Ref.~\cite{lon18} the  proton-charge radius $r_{\text{p}}=0.8751(61)$ fm and the neutron-charge radius $r^2_{\text{n}}=-0.1161(22)$ fm$^2$ recommended by CODATA-2014 were used in the evaluation of nuclear charge radii using Eq.~(\ref{rad_eq}).
Such single-nucleon data come from experiments that study the electron-nucleon system.
 Recently, these quantities were measured more precisely by investigating muonic atoms, and one could ask what is the effect of this increased precision in the nuclear charge radius when applying Eq.~(\ref{rad_eq}).
To address this point in Table~\ref{tab:radii} we compare our results for the charge radii of $^3$He, $^3$H and $^4$He at N2LO using the CODATA-2014 single-nucleon input with the results obtained using the rms proton-charge radius coming from the muonic-hydrogen $r_{\text{p}}=0.84087(39)$~\cite{ant13} and the new value of the rms charge radius of the neutron $r^2_{\text{n}}=-0.106(7)$ fm$^2$~\cite{fil20}. We denote the first choice with $e-r_c$ and the second with  $\mu-r_c$.
The general effect of using this choices of the proton and neutron charge radii amounts to a systematic reduction of roughly 1\% of the charge radii of these light nuclei. This has to be contrasted with the full uncertainty of the theory that includes not only the EIHH numerical error, but also considers  the uncertainty coming from the order-by-order chiral expansion. The latter is estimated using the algorithm proposed first in Ref.~\cite{epe15} and is
included in Table ~\ref{tab:radii}.

\begin{figure}[h!]
\begin{center}
\includegraphics[scale=0.7]{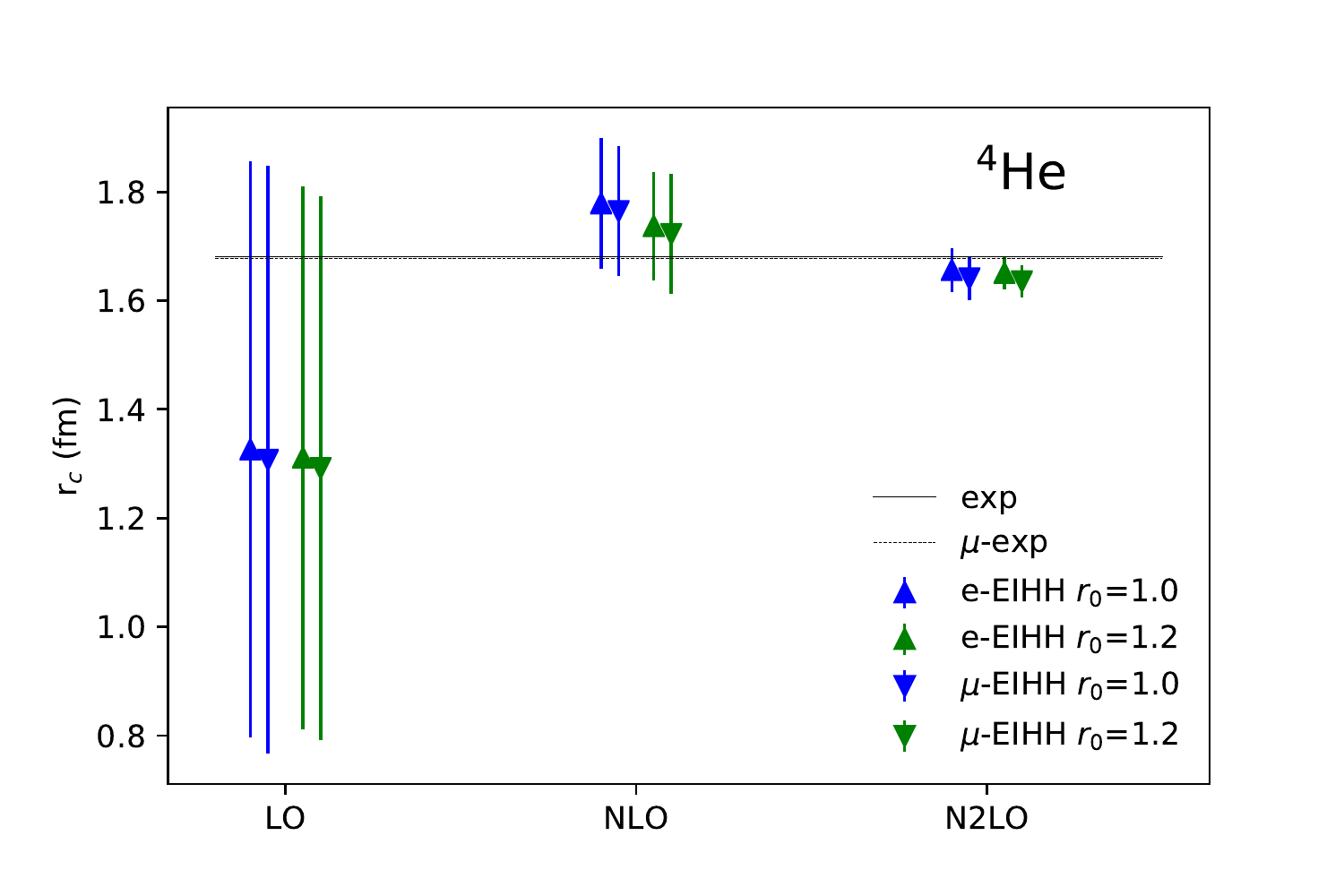}
\end{center}
\caption{The  $^4$He charge radius  computed  at increasing orders of the chiral expansion. The uncertainty bars include the numerical uncertainty of the EIHH method as well as the uncertainties coming from the truncation of the chiral expansion. The horizontal lines are the experimental values from electron scattering (solid line)~\cite{Sick2014} and and from muonic atoms (dashed line)~\cite{kra21}. }
\label{fig:expans}
\end{figure}

For a graphical representation of our findings, in Fig~\ref{fig:expans} we show the $^4$He nuclear charge radius  at increasing chiral orders computed for different choices for the proton and neutron charge radii.
We observe that the chiral order uncertainty  is of the order of 2$\%$, hence larger than the effect of the more precise single-nucleon input.  Overall, we confirm the chiral oder-by-order convergence patter, already discussed in Refs.~\cite{lyn16,lyn17}, but there shown only for the binding energy and the point-proton radius, which does not include the single nucleon input. 

Interestingly, when comparing the $^4$He theoretical charge radius with the newest muonic atom measurement from Ref.~\cite{kra21}, we see that the $\mu-r_c$ results are still consistent with experiment, leaving however space for meson exchange currents to help improving the theoretical precision, which is by far lower than the experimental one.

% Without this important contribution, the precision of our evaluation overcomes the precision coming from the electron-scattering experiments, in some cases by even one order of magnitude, and matches the recent high-precision spectroscopy experiment on the $\mu$-$^4$He atom \cite{kra21}. However, at this level of precision we find out that this N2LO-local interaction tends to under-evaluate the rms charge radius of the $^4$He nucleus, a conclusive statement can be formulated only after the uncertainty coming from the truncation of the chiral order is included. 

\begin{table}[h!t]
\centering
\begin{tabular}{lc ccc c cc cc}
\toprule
 & & \multicolumn{2}{c}{$^3$He} & & \multicolumn{2}{c}{$^3$H} & & \multicolumn{2}{c}{$^4$He}\\
\cline{3-4}
\cline{6-7}
\cline{9-10}

\multirow{2}{*}{} & Cut-off & $\mu$-r$_{\text{c}}$ & $e$-r$_{\text{c}}$ & & $\mu$-r$_{\text{c}}$ & $e$-r$_{\text{c}}$ & & $\mu$-r$_{\text{c}}$ & $e$-r$_{\text{c}}$ \\ & [fm] & [fm] & [fm] & & [fm] & [fm] & & [fm] & [fm] \\
\cline{3-4}
\cline{6-7}
\cline{9-10}
& & & & & & & & & \\
\multirow{2}{*}{EIHH} & 1.0 & 1.96(4) & 1.98(4) & & 1.75(3) & 1.76(3) & & 1.64(4) & 1.66(4) \\ & 1.2 & 1.96(3) & 1.97(3) & & 1.75(3) & 1.76(3) & & 1.64(3) & 1.65(3) \\
& & & & & & & & & \\
Exp, electron & & \multicolumn{2}{c}{1.973(14)} & & \multicolumn{2}{c}{1.759(36)}  & & \multicolumn{2}{c}{1.681(4)} \\
& & & & & & & & & \\
Exp, muon & & \multicolumn{2}{c}{$-$} & & \multicolumn{2}{c}{$-$} & & \multicolumn{2}{c}{1.67824(12)(82)} \\
\bottomrule
\end{tabular}
\caption{Nuclear rms charge radii for $^3$He, $^3$H and $^4$He systems at N2LO computed  using either the single-nucleon  CODATA-2014 values (columns $e$-r$_{\text{c}}$) or the more precise muonic atoms data (columns $\mu$-r$_{\text{c}}$). %For the $\mu$-r$_{\text{c}}$ cases, the neutron charge radius is extracted from the nuclear charge radius obtained in the muonic deuterium experiments, see Ref.~\cite{fil20}.
 The theoretical results are compared to data from the electron-nucleus system~\cite{Sick2014,ANGELI201369} and, when available, to data obtained from the muon-nucleus system~\cite{kra21}.}
\label{tab:radii}
\end{table}

\section{Conclusion and outlook}

In this work, the maximally local chiral interactions
%~\cite{gez14,lyn16,lyn17} 
are implemented for the first time in the hyperspherical harmonic formalism. The benchmark tests performed in light nuclei show general agreement between hyperspherical harmonic results and the previously available Monte Carlo calculations.
%, strengthening the trust in the theoretical derivation and numerical implementation of both methods. 
As expected, at N2LO with the inclusion of the 3N forces  the experimental results are much better reproduced with respect to the LO and NLO calculations. With this study we thus confirm the nice order-by-order convergence in the ground-state energies and in the radii that was already observed in the Monte Carlo studies.

While our numerical precision of the EIHH calculations lies in the sub-percent range, we find that the uncertainties due to the chiral order expansion is higher. In case of the charge radius, we observed that using the most updated values of the proton and nucleon radii instead of the CODATA-2014 values leads to a variation of 1$\%$, which is smaller than the $2\%$ uncertainties found in the chiral order-by-order truncation at N2LO.  Addressing first the latter by going to N3LO should be the priority if the goal is to reduce theoretical uncertainties.
% Finally, the order-by-order convergence of the electric dipole polarizability in $^4$He is studied, the nuclear structure uncertainty is fully estimated by analyzing the truncation of the chiral order. 

Having these new interactions implemented in our formalisms opens up the possibility of investigating other few-body  observables in the future. Our most immediate goals include the investigation of muonic atoms~\cite{muon_review} and of the $^4$He monopole transition form factor~\cite{monopole} in an order-by-order chiral expansions. We reserve these applications to future studies.

%In an upcoming work, the same type of test will be done to muonic atoms systems, due to the enhanced sensitivity of these systems on the nuclear structure, they represent an ideal working frame over which testing commonly accepted state-of-art nuclear interactions. 

\section*{Conflict of Interest Statement}
%All financial, commercial or other relationships that might be perceived by the academic community as representing a potential conflict of interest must be disclosed. If no such relationship exists, authors will be asked to confirm the following statement: 

The authors declare that the research was conducted in the absence of any commercial or financial relationships that could be construed as a potential conflict of interest.

\section*{Author Contributions}

All authors contributed in equal parts to this paper. N.B. derived the first expressions for the spherical tensors,  which were then checked by S.S.L. and S.B.  The two-body force was implemented by S.B. while S.S.L. implemented the expressions of the 3$N$ spherical tensors in the hypershperical harmonics code and run the calculations.
Results were discussed in the group at every step. All authors contributed to the writing of the text.

\section*{Funding}
This work was supported by the
  Deutsche Forschungsgemeinschaft (DFG) through the Collaborative
  Research Center [The Low-Energy Frontier of the Standard Model (SFB
  1044)], and through the Cluster of Excellence ``Precision Physics,
  Fundamental Interactions, and Structure of Matter" (PRISMA$^+$ EXC
  2118/1) funded by the DFG within the German Excellence Strategy
  (Project ID 39083149). Calculations were performed on the mogon2 cluster in Mainz.

\section*{Acknowledgments}
S.S.L. and S.B. would like to acknowledge Joel Lynn, Ingo Tews and Diego Lonardoni for useful discussions.

\bibliographystyle{frontiersinHLTH&FPHY} % for Health, Physics and Mathematics articles
\bibliography{mybib}

\end{document}